\documentclass[pra,singlecolumn,preprintnumbers,superscriptaddress]{revtex4}
\usepackage{bm}
\usepackage{graphicx}
\usepackage{amsbsy}
\usepackage{amsmath}
\usepackage{amsfonts}
\usepackage{amsthm}

\newcommand{\bra}[1]{\langle{#1}|}
\newcommand{\ket}[1]{|{#1}\rangle}

\newcommand{\dg}{^\dagger}

\newcommand{\zo}{\{0,1\}}

\newcommand{\oldx}{\lambda}
\newcommand{\oldn}{r}
\newcommand{\dummy}{k}
\newcommand{\Tau}{m\tau}

\begin{document}

\theoremstyle{plain}
\newtheorem{theorem}{Theorem}
\newtheorem{lemma}{Lemma}
\newtheorem{problem}{Problem}
\newtheorem{corollary}{Corollary}

\title{Efficient quantum algorithms for simulating sparse Hamiltonians}

\author{Dominic W.\ Berry}
\affiliation{Department of Physics, The University of Queensland, Queensland 4072, Australia}
\affiliation{Institute for Quantum Information Science, University of Calgary, Alberta T2N 1N4, Canada}
\author{Graeme Ahokas}
\affiliation{Institute for Quantum Information Science, University of Calgary, Alberta T2N 1N4, Canada}
\affiliation{Department of Computer Science, University of Calgary, Alberta T2N 1N4, Canada}
\author{Richard Cleve}
\affiliation{Institute for Quantum Information Science, University of Calgary, Alberta T2N 1N4, Canada}
\affiliation{Department of Computer Science, University of Calgary, Alberta T2N 1N4, Canada}
\affiliation{School of Computer Science, University of Waterloo, Ontario N2L 3G1, Canada}
\affiliation{Institute for Quantum Computing, University of Waterloo, Ontario N2L 3G1, Canada}
\author{Barry C.\ Sanders}
\affiliation{Institute for Quantum Information Science, University of Calgary, Alberta T2N 1N4, Canada}
\affiliation{Centre for Quantum Computer Technology, Macquarie University, Sydney,
    New South Wales 2109, Australia}

\begin{abstract}
We present an efficient quantum algorithm for simulating the evolution of a
sparse Hamiltonian $H$ for a given time $t$ in terms of a procedure for
computing the matrix entries of $H$. In particular, when $H$ acts on $n$ qubits,
has at most a constant number of nonzero entries in each row/column, and $\|H\|$
is bounded by a constant, we may select any positive integer $k$ such that the
simulation requires $O((\log^*n)t^{1+1/2k})$ accesses to matrix entries of $H$.
We show that the temporal scaling cannot be significantly improved beyond this,
because sublinear time scaling is not possible.
\end{abstract}

\maketitle

\section{Introduction}
There are three main applications of quantum computer algorithms: the hidden
subgroup problem, with Shor's factorization algorithm one important example
\cite{shor}, search problems \cite{grover}, and simulation of quantum systems
\cite{Fey82,lloyd}. Lloyd's method for simulating quantum systems \cite{lloyd}
assumes a tensor product structure of smaller subsystems. Aharonov and Ta-Shma
(ATS) \cite{aha} consider the alternative case where there is no evident tensor
product structure to the Hamiltonian, but it is sparse and there is an efficient
method of calculating the nonzero entries in a given column of the Hamiltonian.
Such respresentations of Hamiltonians can arise as encodings of computational
problems, such as simulations of quantum walks
\cite{walks1,walks2,walks3,walks4,walks5} or adiabatic quantum computations
\cite{adiabatic}.

Here we apply the higher-order integrators of Suzuki \cite{suzuki90,suzuki91}
to reduce the temporal scaling from $t^{3/2}$ \cite{aha} or $t^2$ \cite{lloyd}
to the slightly superlinear scaling $t^{1+1/2k}$, where $k$ is the order of the
integrator and may be an arbitrarily large integer. We determine an upper bound
on the number of exponentials required to approximate the evolution with a given
accuracy. This enables us to estimate the optimal value of $k$, and therefore
the $k$-independent scaling in $t$. We than prove that, in the black-box
setting, this scaling is close to optimal, because it is not possible to perform
simulations sublinear in $t$. We also provide a superior method for decomposing
the Hamiltonian into a sum for the problem considered by ATS, which dramatically reduces the
scaling of $n^2$ \cite{childs} or $n^9$ \cite{aha}
to $\log^*n$ for $n$ qubits. This method is similar to ``deterministic coin
tossing'' \cite{cv}, as well as Linial's graph coloring method \cite{linial}.

\section{Problems and results}
We commence with a statement of the problems that we consider in this paper and
follow with the solutions that will be proven.

\begin{problem}
\label{prob1}
The Hamiltonian is of the form $H = \sum_{j=1}^m H_j$. The problem is to
simulate the evolution $e^{-i H t}$ by a sequence of exponentials
$e^{-i H_j t'}$ such that the maximum error in the final state, as
quantified by the trace distance, does not exceed $\epsilon$. Specifically we wish to
determine an upper bound on the number of exponentials, $N_{\rm exp}$,
required in this sequence.
\end{problem}

For this problem, the $H_j$ should be of a form that permits $e^{-i H_j t'}$ to
be accurately and efficiently simulated for arbitrary evolution time $t'$. It
is therefore reasonable to quantify the complexity of the calculation by the
number of exponentials required. This problem includes the physically
important case of simulating tensor
product systems considered by Lloyd \cite{lloyd}, for which each $H_j$
can be considered to be an interaction
Hamiltonian. It also may be applied to the case where there is a procedure for
calculating the nonzero elements in the columns \cite{aha}. In that case, each
$H_j$ is a 1-sparse Hamiltonian. The decomposition must be calculated, which
requires additional steps in the algorithm.

Our general result for Problem \ref{prob1} is the following theorem.
\begin{theorem}
\label{th1}
When the permissible error is bounded by $\epsilon$, $N_{\rm exp}$
is bounded by
\begin{equation}
\label{Nscaling}
N_{\rm exp} \le 2m5^{2k} (\Tau)^{1+1/2k}/\epsilon^{1/2k},
\end{equation}
for $\epsilon\le 1 \le 2m 5^{k-1}\tau$, where $\tau=\|H\|t$, and $k$ is
an arbitrary positive integer.
\end{theorem}

By taking $k$ to be sufficiently large, it is possible to obtain
scaling that is arbitrarily close to linear in $\tau$. However,
for a given value of $\tau$, taking $k$ to be too large will
increase $N_{\rm exp}$. To estimate the optimum value of $k$ to take, we
express Eq.~(\ref{Nscaling}) as
\[
N_{\rm exp} \le 2m^2 \tau\, e^{2k\ln 5 + \ln (m\tau/\epsilon)/2k}.
\]
The right-hand side has a minimum for
\[
k={\rm round}\left[ \frac 12 \sqrt{\log_5 (m\tau/\epsilon)+1} \right].
\]
Here we have added 1 and rounded because~$k$ must take integer
values. Adopting this value of $k$ provides the upper bound
\begin{equation}
    N_{\rm exp} \le 4m^2\tau\, e^{2\sqrt{\ln 5 \ln (m\tau/\epsilon)}} ,
\label{eq:Nexpwithoutk}
\end{equation}
for $\epsilon\le 1 \le m\tau/25$. Eq.~(\ref{eq:Nexpwithoutk})
is an expression for $N_{\rm exp}$ that is independent of~$k$.

The scaling in Eq.~(\ref{eq:Nexpwithoutk})
is close to linear for large $m\tau$. We show that this scaling is
effectively optimal, because it is not possible to perform general simulations
sublinear in $\tau$ (see Sec.\ \ref{sec:lin}). This result applies in the
``black-box'' setting, so it does not rule out the possibility that individual
Hamiltonians have structure which allows them to be simulated more efficiently.

The second problem which we consider is that of sparse Hamiltonians.
\begin{problem}
\label{prob2}
The Hamiltonian $H$ has no more than $d$ nonzero entries in each column, and
there exists a black-box function $f$ that gives these entries. The dimension
of the space which $H$ acts upon does not exceed $2^n$. If the nonzero
elements in column~$x$ are $y_1,\ldots,y_{d'}$, where $d'\le d$, then $f(x,i)=
(y_i,H_{x,y_i})$ for $i\le d'$, and $f(x,i)=(x,0)$ for $i>d'$. The problem
is to simulate the evolution $e^{-i H t}$ such that the maximum error in the
final state, as quantified by the trace distance, does not exceed $\epsilon$.
We wish to determine the scaling of the number of calls to $f$, $N_{\rm bb}$,
required for this simulation.
\end{problem}

For each~$x$, the order in which the corresponding $y_i$ are given can be
arbitrary. The function $f$ is an arbitrary black-box function, but we assume
that there is a corresponding unitary $U_f$ such that
\[ U_f \ket{x,i}\ket 0=\ket{\phi_{x,i}}\ket{y_i,H_{x,y_i}}, \]
and we may perform calls to both $U_f$ and $U_f\dg$. Here $\ket{\phi_{x,i}}$
represents any additional states which are produced in the reversible
calculation of $f$.

ATS approached the problem by decomposing the Hamiltonian into a sum of $H_j$.
We apply a similar approach to obtain the following theorem.
\begin{theorem}
\label{prop2}
The number of black-box calls for given $k$ is
\begin{equation}
\label{calls}
N_{\rm bb} \in O\left((\log^*n) d^2 5^{2k}
(d^2\tau)^{1+1/2k}/\epsilon^{1/2k}\right)
\end{equation}
with $\log^*n\equiv\min\{ r | \log_2^{(r)} n<2\}$ (the $^{(r)}$ indicating
the iterated logarithm).
\end{theorem}
The $\log^*n$ scaling is a dramatic improvement over the $n^9$ scaling implicit in
the method of ATS, as well as the $n^2$ scaling of Childs \cite{childs}.

\section{Higher order integrators}
To prove Theorem~\ref{th1}, we apply the method of higher-order integrators.
Following Suzuki \cite{suzuki90,suzuki91}, we define
\[
S_2(\oldx)=\prod_{j=1}^m e^{H_j\oldx/2}\prod_{j'=m}^1 e^{H_{j'}\oldx/2},
\]
which is the basic Lie-Trotter product formula, and the recursion relation
\[
S_{2k}(\oldx) = [S_{2k-2}(p_k \oldx)]^2 S_{2k-2}((1-4p_k)\oldx)
[S_{2k-2}(p_k \oldx)]^2
\]
with $p_k=(4-4^{1/(2k-1)})^{-1}$ for $k>1$. Suzuki then proves that \cite{suzuki90}
\begin{equation}
\label{suz1}
    \left\| \exp\left(\sum_{j=1}^m H_j \oldx \right) - S_{2k}(\oldx)\right\| \in O(|\oldx|^{2k+1})
\end{equation}
for $|\oldx| \to 0$. The parameter $\oldx$ corresponds to $-it$ for Hamiltonian
evolution.

We first assess the higher-order integrator method in terms of all quantities
$t$, $m$, $k$, and $\|H\|$. Our result is
\begin{lemma}
\label{prop:integrator_k}
Using integrators of order $k$ and dividing the time into~$r$ intervals, we have
the bound
\begin{equation}
\label{new2}
\left\| \exp\left( -it\sum_{j=1}^m H_j \right)-
    [S_{2k}(-it/\oldn)]^\oldn \right\|
 \le 5(2\times 5^{k-1} m \tau)^{2k+1}/\oldn^{2k},
\end{equation}
for
\begin{align}
\label{restrict}
4m5^{k-1}\tau/\oldn &\le 1, \nonumber \\
(16/3)(2\times 5^{k-1}m\tau)^{2k+1}/\oldn^{2k}&\le 1.
\end{align}
\end{lemma}
\begin{proof}
Consider a Taylor expansion of both terms in the left-hand side (LHS) of
Eq.~(\ref{suz1}). Those terms containing $\oldx$ to powers less than $2k+1$
must cancel because the correction term is $O(|\oldx|^{2k+1})$, and terms with
$\oldx^{2k'+1}$ for $k'\ge k$ must contain a product of $2k'+1$ of the $H_j$
terms; thus
\[
    \exp\left(\sum_{j=1}^m H_j \oldx\right) = S_{2k}(\oldx)
        +\sum_{k'=k}^{\infty} \oldx^{2k'+1}\sum_{l=1}^{L_{k'}} C_l^{k'}
            \prod_{q=1}^{2k'+1}H_{j_{lq}} .
\]
The constants $C_l^{k'}$ and the number of terms $L_{k'}$ depend on~$m$ and~$k$.

In order to bound $C_l^{k'}$ and $L_{k'}$, first consider the Taylor expansion
of the exponential in the LHS of Eq.~(\ref{suz1}). Because the operators~$H_j$ are
in general noncommuting, expanding $(H_1+\cdots+H_m)^{2k'+1}$ yields $m^{2k'+1}$
terms. Therefore the Taylor expansion contains $m^{2k'+1}$ terms with
$\oldx^{2k'+1}$. These terms have multiplying factors of $1/(2k'+1)!$ because
this is the multiplying factor given by the Taylor expansion of the exponential.

To place a bound on the number of terms in the Taylor expansion of $S_{2k}(\oldx)$,
note that $S_{2k}(\oldx)$ consists of a product of $$2(m-1)5^{k-1}+1$$
exponentials. The expansion for $S_{2k}(t)$ may be obtained by expanding each of
the exponentials individually. There will be no more than
$$[2(m-1)5^{k-1}+1]^{2k'+1}$$ terms with $\oldx^{2k'+1}$. Because $|p_k|<1$ and
$|1-4p_k|<1$, the multiplying factors for each of these terms must be less than
1.

Each $H_j$ satisfies $\|H_j\|\le\| H \|$ \cite{aha}. Defining
$\Lambda\equiv\| H \|$, and using standard inequalities we obtain
\begin{align}
    &\left\| \sum_{k'=k}^{\infty}\oldx^{2k'+1}\sum_{l=1}^{L_{k'}} C_{l}^{k'}
    \prod_{q=1}^{2k'+1}H_{j_{lq}} \right\| \le
    \sum_{k'=k}^{\infty} |\oldx\Lambda|^{2k'+1}L_{k'} \nonumber \\ &\le
    \sum_{k'=k}^{\infty} |\oldx\Lambda |^{2k'+1}\big\{ m^{2k'+1}  +[2(m-1)5^{k-1} +1]^{2k'+1} \big\} \nonumber \\ &\le
    2\sum_{k'=k}^{\infty} |\oldx\Lambda |^{2k'+1}
    [2m 5^{k-1}]^{2k'+1} = \frac{2|2m 5^{k-1}\oldx\Lambda|^{2k+1}}{1-|2m 5^{k-1}\oldx\Lambda|^2}. \nonumber
\end{align}

Therefore we obtain the inequality
\[
    \left\| \exp\left( \oldx \sum_{j=1}^m H_j \right) - S_{2k}(\oldx)\right\|
    \le (8/3)|2m 5^{k-1} \Lambda \oldx|^{2k+1},
\]
provided $|2m 5^{k-1}\Lambda \oldx| \le 1/2$. Substituting $\lambda=-it/r$ where
$r$ is an integer, and taking the power of $r$, gives the error bound
\begin{equation}
\label{newap}
    \left\| \exp\left( -it\sum_{j=1}^m H_j \right)
        -[S_{2k}(-it/\oldn)]^\oldn \right\| \le  [1+(8/3)(2m 5^{k-1}\Lambda t/\oldn)^{2k+1}]^\oldn-1,
\end{equation}
for $4m 5^{k-1}\Lambda t/\oldn \le 1$.
This may alternatively be expressed as in Lemma \ref{prop:integrator_k}.
\end{proof}

By placing limits on the norm of the difference in the unitaries, we limit the
trace distance of the output states. This is because
\begin{align}
\|U_1-U_2\|&\ge \|U_1\ket{\psi}-U_2\ket{\psi}\| \nonumber \\ &\ge
\frac 12 {\rm Tr}\left|U_1\ket{\psi}\bra{\psi}U_1\dg-U_2\ket{\psi}\bra{\psi}U_2\dg\right| \nonumber \\
&= D\left( U_1\ket{\psi}\bra{\psi}U_1\dg,U_2\ket{\psi}\bra{\psi}U_2\right), \nonumber
\end{align}
with~$D$ the trace distance. We now use this to prove Theorem~\ref{th1}.

\begin{proof} \emph{(of Theorem~\ref{th1})}
Let us take
\begin{equation}
\label{rval}
r = \lceil 4\times 5^{k-1/2}(\Tau)^{1+1/2k}/\epsilon^{1/2k}\rceil .
\end{equation}
Given the restrictions $\epsilon\le 1 \le 2m 5^{k-1}\tau$,
it is easily seen that Eqs.~(\ref{restrict}) hold.
In addition, the right-hand side of Eq.~(\ref{new2}) does not exceed $\epsilon$, so the error
can not exceed $\epsilon$.

Because the number of exponentials in $S_{2k}(\lambda)$ does not exceed
$2m5^{k-1}$, we have $N_{\rm exp} \le 2m5^{k-1}r$. If we take $r$ as in Eq.~(\ref{rval}), then
we find that
\begin{equation}
\label{Nineq}
N_{\rm exp} \le 2m5^{2k} (\Tau)^{1+1/2k}/\epsilon^{1/2k}.
\end{equation}
Here the multiplying factor has been changed to take into account the
ceiling function. Hence the order scaling is as in Eq.~(\ref{Nscaling}).
\end{proof}

This result may be used for any case where the Hamiltonian is a sum of
terms that may be simulated efficiently.  It may therefore be applied to the
case of tensor product systems, where the individual $H_j$ are interaction
Hamiltonians. It can be also used for cases of the type of Problem~2,
where the Hamiltonian is sparse. In this case we have the additional task of
decomposing the Hamiltonian into a sum.

\section{Linear limit on simulation time}
\label{sec:lin}
We have shown that the simulation of any Hamiltonian may be performed
arbitrarily close to linearly in the scaled time $\tau$.
We now show that the scaling cannot be sublinear in $\tau$, provided the
number of qubits can grow at least logarithmically with respect to $\tau$.
The result is
\begin{theorem}
\label{th3}
For all positive integers $N$ there exists a row-computable
$2$-sparse Hamiltonian $H$ such that simulating the evolution of $H$ for scaled
time $\tau = \pi N/2$ within precision $1/4$ requires at least
$\tau/2\pi$ queries to $H$.
\end{theorem}

Here a row-computable Hamiltonian means one where there is a method for
efficiently calculating the nonzero elements in each row.
\begin{proof}
The idea is to construct a 2-sparse Hamiltonian such that the simulation of
this Hamiltonian determines the parity of $N$ bits. It has been shown that the
parity of $N$ bits requires $N/2$ queries to compute within error $1/4$
\cite{BealsEtAl}; therefore the Hamiltonian can not be simulated any more
efficiently.

First consider a Hamiltonian $H$ acting on orthogonal basis states $\ket{0},
\ldots, \ket{N}$, for which the nonzero matrix entries are
\[ \bra{j+1}H\ket{j}= \bra{j}H\ket{j+1} =\sqrt{(N-j)(j+1)}/2. \]
This Hamiltonian is equivalent to
a $J_x$ operator for a spin $N/2$ system, with the $\ket{j}$ being $J_z$
eigenstates. It is therefore clear that $e^{-i\pi H}\ket{0} = \ket{N}$ and
$\|H\|=N/2$.

Now we construct an augmented version of the above Hamiltonian, that corresponds
to a graph with two disjoint lines with weights as above, where the lines
``cross over'' at the positions where bits $X_1,\ldots, X_N$ are 1. We add an
ancilla qubit so the Hamiltonian~$H$ acts on basis states $\ket{0,0}, \ldots,
\ket{0,N}$, $\ket{1,0}, \ldots, \ket{1,N}$. The nonzero matrix entries of $H$
are
\[
\bra{k',j+1}H\ket{k,j} = \bra{k,j}H\ket{k',j+1} = \sqrt{(N-j)(j+1)}/2
\]
for values of~$k$ and~$k'$ such that $k \oplus k' = X_{j+1}$ (where $\oplus$ is XOR).

Thus, if $X_{j+1}$ is zero, then there is a nonzero matrix element between
$\ket{0,j}$ and $\ket{0,j+1}$, as well as between $\ket{1,j}$ and $\ket{1,j+1}$.
If $X_{j+1}$ is equal to 1, then the nonzero matrix elements are between
$\ket{0,j}$ and $\ket{1,j+1}$, as well as $\ket{1,j}$ and $\ket{0,j+1}$.
We may determine a sequence of bits $k_0,\ldots,k_N$ such that $k_j\oplus
k_{j+1}=X_{j+1}$. The Hamiltonian acting on the set of states $\ket{k_j,j}$
will then be identical to that acting on the states $\ket{j}$ with the original
Hamiltonian. It is therefore clear that $e^{-i\pi H}\ket{k_0,0} = \ket{k_N,N}$.

\begin{figure}
\centerline{\includegraphics[width=9cm]{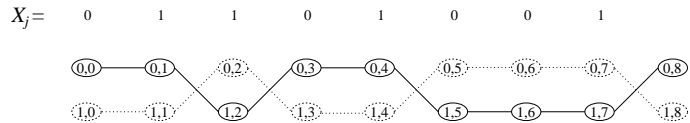}}
\caption{Graph representing example the Hamiltonian in the proof of Theorem
\ref{th3}. States are represented by ellipses, and nonzero elements of the
Hamiltonian are indicated by lines. The sequence of states $\ket{k_j,j}$ with
$k_0=0$ is indicated by the solid line.}
\label{fig}
\end{figure}

The graph corresponding to a Hamiltonian of this type is shown in Fig.~\ref{fig}.
The system separates into two distinct sets of states which are not connected.
If the system starts in one of the states on the path indicated by the solid
line, it can not evolve under the Hamiltonian to a state on the dotted line.
{}From the definition of the $k_j$, if $k_0=0$, then $k_j$ is the parity of bits
$X_1$ to $X_j$, and in particular $k_N$ gives the parity of all $N$ bits.
Thus if we start with the initial state $\ket{0,0}$ and simulate the evolution
$e^{-i\pi H}$, we obtain the state $\ket{k_N,N}$, where $k_N$ is the parity of
the $N$ bits $X_1,\ldots,X_N$. Thus measuring the state of the ancilla qubit
will give the parity.

Let us denote the final state obtained by the simulation by $\ket\psi$, and the
reduced density operator for the ancilla by $\rho_{\rm anc}$. If the error
probability is no less than $1/4$, then $D(\rho_{\rm anc},\ket{k_N}\bra{k_N})
\ge 1/4$, which implies that
\[ D(\ket{\psi}\bra{\psi},\ket{k_N,N}\bra{k_N,N})\ge 1/4. \]
Hence, if there are no more than $N/2$ queries to the $X_j$, the
error in the simulation as quantified by the trace distance must be at least
$1/4$.

Each query to a column of $H$ requires no more than two queries to the $X_j$
(for column~$j$ we require a query to $X_j$ and $X_{j+1}$). Thus, if there are
no more than $N/4$ queries to $H$, then there are no more than $N/2$ queries to
the $X_j$. In addition, the scaled time for the simulation is
\[  \tau=\|H\|t=\pi N/2. \]
Thus the simulation of $H$ requires at least $N/4 = \tau/2\pi$ queries
to obtain trace distance error no more than $1/4$.
\end{proof}

The form of this result differs slightly from that in the previous section, in
that the cost is specified in terms of the number of queries to the Hamiltonian,
rather than the number of exponentials. It is straightforward to show the
following result for the number of exponentials.

\begin{corollary}
There is no general integrator for Hamiltonians of the form $H=H_1+H_2$ such
that (trace distance) error $\le 1/4$ may be achieved with the number of
exponentials $N_{\rm exp}<\tau/2\pi$.
\end{corollary}

By general integrator we mean an integrator that depends only on $\tau$, and
not the Hamiltonian.
\begin{proof}
We take $H$ as in the preceding proof. This Hamiltonian may be expressed in the
form $H=H_1+H_2$ by taking $H_1$ to be the Hamiltonian with
$\bra{k',j+1}H_1\ket{k,j}$ nonzero only for even~$j$, and $H_2$ to be the
Hamiltonian with $\bra{k',j+1}H_2\ket{k,j}$ nonzero only for odd~$j$. Each query
to the $H_k$ requires only one query to the $X_j$. For $H_1$ ($H_2$),
determining the nonzero element in column~$j$ requires determining which of~$j$
and $j+1$ is odd (even), and performing a query to the corresponding $X_j$ or
$X_{j+1}$.

Both $H_1$ and $H_2$ are 1-sparse, and therefore may be efficiently simulated
with only two queries \cite{ccdfgs,ahokas}. If $N_{\rm exp}<\tau/2\pi$, the
total number of queries to the $X_j$ is no more than $\tau/\pi$. Taking $t=\pi$
and $\| H \|=N/2$, the number of queries is no more than $N/2$. However, from
Ref.\ \cite{BealsEtAl} the error rate can be no less than $1/4$. Hence error
rate $\le 1/4$ can not be achieved with $N_{\rm exp}<\tau/2\pi$.
\end{proof}

\section{Efficient decomposition of Hamiltonian}
Next we consider the problem of simulating general sparse Hamiltonians, as in
Problem \ref{prob2}. Given that the dimension of the space does not exceed
$2^n$, we may represent the state of the system on~$n$ qubits, and~$x$ and $y$
may be $n$-bit integers. The real and imaginary parts of the matrix elements
will be represented by $n'$ bit integers (for a total of $2n'$ bits for each
matrix element), where $n'$ must be chosen large enough to achieve the desired
accuracy. 

In order to simulate the Hamiltonian, we decompose it into the form
$H = \sum_{j=1}^m H_j$, where each $H_j$ is \textit{1-sparse} (i.e., has at
most one nonzero entry in each row/column). If $H_j$ is 1-sparse then it is
possible to directly simulate $\exp(-i H_j t)$ with just two black-box
queries to $H_j$ \cite{ccdfgs,ahokas}. Since the value of $m$ directly impacts
the total cost of simulating $H$, it is desirable to make $m$ as small as
possible. The size of the sum may be limited as in the following lemma.
\begin{lemma}
\label{declem}
    There exists a decomposition $H = \sum_{j=1}^m H_j$, where each $H_j$ is
    $1$-sparse, such that $m=6d^2$
    and each query to any $H_j$
    can be simulated by making $O(\log^{\ast} n)$ queries to $H$.
\end{lemma}

\begin{proof}
{}From the black-box function for $H$, we wish to determine black-box functions
for each $H_j$ that give the nonzero row number, $y$, and matrix element
corresponding to each column~$x$. This black-box for $H_j$ is represented by
the function $g(x,j)$, with output $(y,(H_j)_{x,y})$.
If there is no nonzero element in column~$x$, the output is $(x,0)$.

Intuitively, it is helpful to consider the graph $G_H$
associated with $H$ whose vertex set is $\zo^n$. Each vertex corresponds to a
row or column number, and there is an edge between vertex~$x$ and $y$ if
the matrix element $H_{x,y}$ is nonzero. As $H$ is Hermitian we take the
graph to be undirected. We wish to determine an ``edge-coloring'' of $G_H$,
which is a labeling of the edges such that incident edges have different
colors. Each edge color,~$j$, then corresponds to a different Hamiltonian
$H_j$ in the decomposition of $H$.

The basic idea is as in the following labeling scheme, where the labels are indexed from
the set $\{1,\ldots,d\}^2$. We take $f_y$ to be the $y$-component of $f$; then
$f_y(x,i)$ gives the $i^\text{th}$ neighbor of vertex~$x$ in the graph. Let $(x,y)$ be
an edge of $G_H$ such that $y=f_y(x,i)$ and $x=f_y(y,j)$. Thus edge~$(x,y)$ is
labeled with the ordered pair~$(i,j)$ for $x \le y$,
or $(j,i)$ for $x\ge y$. This labeling is not quite an edge-coloring;
for $w<x<y$ it is possible for edges $(w,x)$ and $(x,y)$
to both have the label $(i,j)$. That will be the case if $y$ and~$w$ are the
$i^\text{th}$ and $j^\text{th}$ neighbors of~$x$, respectively, and~$x$ is the $i^\text{th}$
neighbor of~$w$ and the $j^\text{th}$ neighbor of $y$. To ensure that the labels are unique, we
add the additional parameter~$\nu$, so the label is $(i,j,\nu)$.

We assign $\nu$ via a method similar to ``deterministic coin tossing''
\cite{cv}. We set $x_0^{(0)}=x$, then determine a sequence of vertices
$$x_0^{(0)}<x_1^{(0)}<x_2^{(0)}<\cdots$$ such that $x_{l+1}^{(0)}=
f_y(x_l^{(0)},i)$ and $f_y(x_{l+1}^{(0)},j)=x_l^{(0)}$. That is, the edges
$(x_l^{(0)}$, $x_{l+1}^{(0)})$ are labeled $(i,j,\nu)$, with the same values of
$i$ and~$j$ for each edge. We need to choose values of $\nu$ for the edges
such that the same value is never repeated in this chain.

A typical chain may have only two elements; however, there
exist Hamiltonians such that long chains may be formed. In the case that the
chain is long, we do not determine it any further than $x_{z_n+1}^{(0)}$. Here
$z_n$ is the number of times we must iterate $l \mapsto 2\lceil \log_2 l \rceil$
(starting at $2^n$) to obtain 6 or less. This quantity is of order $\log^*n$, and for
any realistic problem size $z_n$ itself will be no more than 6 \footnote{For $z_n>6$
we require $n>10^{10^{37}}$; clearly an unrealistic problem size.}.

Now we determine a second sequence of values $x_l^{(1)}$. This sequence is taken
to have the same length as the first sequence.
For each $x_l^{(0)}$ and $x_{l+1}^{(0)}$, we determine the first bit position
where these two numbers differ, and record the value of this bit for
$x_l^{(0)}$, followed by the binary representation of this position, as
$x_l^{(1)}$. The bit positions are numbered from zero; that is, the first bit is
numbered $00\ldots 0$. If $x_l^{(0)}$ is at the end of the sequence, we simply
take $x_l^{(1)}$ to be the first bit of $x_l^{(0)}$, followed by the binary
representation of 0. There are $2^n$ different possible values for each of the
$x_l^{(0)}$, and $2n$ different possible values for each of the $x_l^{(1)}$.

From the definition, each~$x_l^{(0)}$ is unique. Also $x_l^{(1)}$ must
differ from $x_{l+1}^{(1)}$. This is because, even if the positions of the first
bit where $x_l^{(0)}$ differs from $x_{l+1}^{(0)}$ and $x_{l+1}^{(0)}$ differs
from $x_{l+2}^{(0)}$ are identical, the value of this bit for $x_l^{(0)}$ will
of course be different from the value for $x_{l+1}^{(0)}$. As the $x_l^{(1)}$
contain both the position and the value of the bit, $x_l^{(1)}$ must differ
from $x_{l+1}^{(1)}$.

There is a subtlety when $x_{l+1}^{(0)}$ is at the end of the sequence. Then
$x_{l+1}^{(1)}$ contains the first bit of $x_{l+1}^{(0)}$, and the position of
the first bit which differs is taken to be 1. In that case, if $x_{l}^{(0)}$
differs from $x_{l+1}^{(0)}$ at the first bit (so the bit positions recorded in
$x_{l}^{(1)}$ and $x_{l+1}^{(1)}$ are identical), then the bit values which are
recorded in $x_{l}^{(1)}$ and $x_{l+1}^{(1)}$ must be different. Thus it is
still not possible for $x_{l}^{(1)}$ to be equal to $x_{l+1}^{(1)}$.

We repeat this process until we determine the sequence of values
$x_l^{(z_n)}$. We determine the $x_l^{(p+1)}$ from the
$x_l^{(p)}$ in exactly the same way as above. At each step, $x_l^{(p)}$ differs
from $x_{l+1}^{(p)}$ for exactly the same reasons as for $p=1$.
As we go from $p$ to $p+1$, the number of possible values for the $x_l^{(p)}$
is reduced via the mapping $\dummy \mapsto 2\lceil \log_2 \dummy \rceil$.
Due to our choice of $z_n$, there are six possible values for $x_{0}^{(z_n)}$.

Now if $w<x$ with $x=f_y(w,i)$ and $w=f_y(x,j)$, then we may set $w_0^{(0)}=w$
and perform the calculation in exactly the same way as for~$x$ in order to
determine $w_{0}^{(z_n)}$. If the chain of $x_{l}^{(0)}$ ends before $z_n$,
then the $x_l^{(p)}$ will be the same as the $w_{l+1}^{(p)}$. In particular
$x_0^{(z_n)}$ will be equal to $w_1^{(z_n)}$, so it is clear that $w_0^{(z_n)}$
will differ from $x_0^{(z_n)}$.

On the other hand, if there is a full chain of $x_{0}^{(0)}$ up to
$x_{z_n+1}^{(0)}$, then the chain for~$w$ will end at $w_{z_n+1}^{(0)}$, which
is equivalent to $x_{z_n}^{(0)}$. Then $w_{z_n+1}^{(1)}$ will be calculated in
a different way to $x_{z_n}^{(1)}$, and may differ. However, $w_{z_n}^{(1)}$
will be equal to $x_{z_n-1}^{(1)}$. At step $p$, $w_{z_n-p+1}^{(p)}$ will be
equal to $x_{z_n-p}^{(p)}$. In particular, at the last step, $w_{1}^{(z_n)}$
will be equal to $x_{0}^{(z_n)}$. Thus we find that $w_{0}^{(z_n)}$ again
differs from $x_{0}^{(z_n)}$.

As $x_{0}^{(z_n)}$ has this useful property, we assign the edge $(x,y)$ the
color $(i,j,\nu)$, where $\nu=x_{0}^{(z_n)}$. Due to the properties of the
above scheme, if the edge $(w,x)$ has the same values of~$i$ and~$j$ as
$(x,y)$, it must have a different value of $\nu$. Therefore, via this scheme,
adjacent edges must have different colors.

Now we describe how to calculate the black-box function $g$ using this
approach. We replace~$j$ with $(i,j,\nu)$ to reflect the labeling
scheme, so the individual Hamiltonians are $H_{(i,j,\nu)}$. The black-box
function we wish to calculate is $g(x,i,j,\nu)$. We also define the function
$\Upsilon(x,i,j)$ to be equal to the index $\nu$ as calculated in the above
way. There are three main cases where we give a nontrivial output:
\begin{enumerate}
    \item $f_y(x,i)=x$, $i=j$ and $\nu=0$,
    \item $f_y(x,i)>x$, $f_y(f_y(x,i),j)=x$ and $\Upsilon(x,i,j)=\nu$,
    \item $f_y(x,j)<x$, $f_y(f_y(x,j),i)=x$ and $\Upsilon(f_y(x,j),i,j)=\nu$.
\end{enumerate}
In cases 1 and 2 we return $g(x,i,j,\nu)=f(x,i)$, and for case 3 we return
$g(x,i,j,\nu)=f(x,j)$; in all other cases we return $g(x,i,j,\nu)=(x,0)$.

Case~1 corresponds to diagonal elements of the Hamiltonian. We only return a
nonzero result for $\nu=0$, in order to prevent this element being repeated in
different Hamiltonians $H_{(i,j,\nu)}$. Case~2 corresponds to there existing a
$y>x$ such that $y$ is the $i^\text{th}$ neighbor of~$x$ and~$x$ is the $j^\text{th}$
neighbor of $y$. Similarly Case~3 corresponds to there existing a
$w<x$ such that~$w$ is the $j^\text{th}$ neighbor of~$x$ and~$x$ is the $i^\text{th}$
neighbor of~$w$. The uniqueness of the labeling ensures that cases 2 and 3 are
mutually exclusive.

As there are $d$ possible values for~$i$ and~$j$, and $\nu$ may take six values,
there are $6d^2$ colors. Thus we may take $m=6d^2$. In determining $\nu$,
we need a maximum of $2(z_n+2)$ queries to the black-box; this is of order
$\log^* n$. \end{proof}

\begin{table*}
\caption{Example values of~$x_l^{(p)}$ under our scheme for calculating $\nu$.
The value of $\nu$ obtained is in the upper right, and is shown in bold.
For this example~$n=18$ and~$z_n=4$. The values in italics are those that may
differ from~$w_{l+1}^{(p)}$ (there are no corresponding values for the bottom
row).
\label{xtable}} \centering{\begin{tabular}{|c|c|c|c|c|c|c|c|c|} \hline
$l\backslash p$ & 0 & 1 & 2 & 3 & 4 \\ \hline
0 & 001011100110011010 & 000001 & 0100 & 000 & \textbf{000} \\
1 & 010110101010011011 & 000010 & 1100 & 100 & \textit{100} \\
2 & 011011101110101101 & 000000 & 0001 & \textit{000} & \textit{000} \\
3 & 101011101011110100 & 010001 & \textit{1001} & \textit{100} & \textit{100} \\
4 & 101011101011110101 & \textit{000001} & \textit{0000} & \textit{000} & \textit{000} \\
5 & 111000010110011010 & 100000 & 1000 & 100 & 100 \\ \hline
\end{tabular}}
\end{table*}

\begin{table*}
\caption{Example values of $w_l^{(p)}$ under our scheme for calculating $\nu$.
The value of $\nu$ obtained is in the upper right, and is shown in bold. For
this example $n=18$ and $z_n=4$. The values in italics are those which may
differ from $x_{l-1}^{(p)}$.
\label{wtable}} \centering{\begin{tabular}{|c|c|c|c|c|c|c|c|c|} \hline
$l\backslash p$ & 0 & 1 & 2 & 3 & 4 \\ \hline
0 & 000010010110111001 & 000010 & 1100 & 100 & \textbf{100} \\
1 & 001011100110011010 & 000001 & 0100 & 000 & 000 \\
2 & 010110101010011011 & 000010 & 1100 & 100 & \textit{001} \\
3 & 011011101110101101 & 000000 & 0001 & \textit{111} & \textit{100} \\
4 & 101011101011110100 & 010001 & \textit{0000} & \textit{000} & \textit{000} \\
5 & 101011101011110101 & \textit{100000} & \textit{1000} & \textit{100} & \textit{100} \\ \hline
\end{tabular}}
\end{table*}

\begin{figure}
\centerline{\includegraphics[width=9cm]{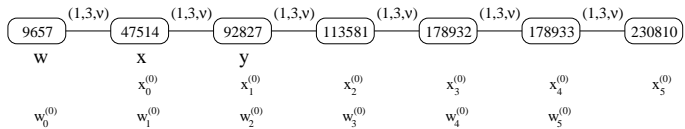}}
\caption{A portion of the graph for the example given in Tables \ref{xtable} and
\ref{wtable}. The vertices~$w$,~$x$, $y$, etc each have $i=1$ and $j=3$ for the
edge labels, so it is necessary for the $\nu$ to differ to ensure that adjoining
edges have distinct labels. The $x_l^{(0)}$ and $w_l^{(0)}$ which the vertices
correspond to are also given. The numbers in the first columns of Tables
\ref{xtable} and \ref{wtable} are the binary representations of the vertex
numbers given here.}
\label{graph}
\end{figure}

To illustrate the method for determining $\nu$, an example is given in Table~\ref{xtable}
for~$\{x_l^{(p)}\}$ and Table~\ref{wtable} for~$\{w_l^{(p)}\}$.
Fig.~\ref{graph} shows a portion of the graph corresponding to the values
in Tables~\ref{xtable} and~\ref{wtable}. In the Tables $n=18$, so
there are $2^{18}$ possible values in the first column. Then there are 36 possible
values in the second column, 12 in the third, 8 in the fourth and 6 in the fifth.
Thus $z_n$ is equal to 4 in this case, and the sequence of
$x_l^{(0)}$ is determined up to $x_5^{(0)}$.

In Table \ref{xtable} the values of $x_l^{(p)}$ are given, and the elements in
the first column are values of $x_l^{(0)}$. As an example of calculation of
$x_l^{(1)}$, note that $x_0^{(0)}$ differs from $x_1^{(0)}$ in the second bit
position. The second bit for $x_0^{(0)}$ is 0, so this is the first bit for
$x_0^{(1)}$. We subtract 1 from the bit position to obtain 1, and take the
remaining bits of $x_0^{(1)}$ to be the binary representation of 1. For the case
of $x_5^{(0)}$, this is the end of the chain, so we simply take $x_5^{(1)}$ to
be the first bit of $x_5^{(0)}$, which is 1, and the binary representation of 0.

In Table \ref{wtable}
the values of $w_l^{(p)}$ are given, where these are calculated from a $w<x$
such that $x=f_y(w,i)$ and $w=f_y(x,j)$. The example given illustrates the case
where the sequence of $w_l^{(0)}$ (with $w_l^{(0)}=x_{l-1}^{(0)}$) ends before
the sequence of $x_l^{(0)}$. In this case, we find that the differences
propagate towards the left, but we still have $x_{0}^{(z_n)}=w_{1}^{(z_n)}$.
Thus different values of $\nu$ are obtained, as expected. For~$x$ we obtain
$\nu=x_0^{(4)}=000$, and for~$w$ we obtain $\nu=w_0^{(4)}=100$.

We can use Lemma \ref{declem} to prove Theorem~\ref{prop2}.
\begin{proof} {\it (of Theorem~\ref{prop2})}
Overall the number of Hamiltonians $H_{(i,j,\nu)}$ in the decomposition is
$m=6d^2$. To calculate $g(x,i,j,\nu)$, it is necessary to call the
black-box $2(z_n+2)$ times.

To simulate evolution under the Hamiltonian $H_{(i,j,\nu)}$, we require
$g$ to be implemented by a unitary operator $U_g$ satisfying
\[ U_g \ket{x,i,j,\nu}\ket 0=
\ket{x,i,j,\nu}\ket{y,(H_{i,j,\nu})_{x,y}}. \]
As discussed above, the function $f$ may be represented by a unitary $U_f$;
using this unitary it is straightforward to obtain a unitary $\tilde U_g$
such that
\[ \tilde U_g \ket{x,i,j,\nu}\ket 0=
\ket{\phi_{x,i,j,\nu}}\ket{y,(H_{i,j,\nu})_{x,y}}. \]
We may obtain the unitary $U_g$ in the usual way by applying $\tilde U_g$,
copying the output, then applying $\tilde U_g\dg$ \cite{mikeike}.

Using the method of Ref.\ \cite{aha}, the Hamiltonian $H_{(i,j,\nu)}$ may be
simulated using a call to $U_g$ and a call to $U_g\dg$. As $z_n$ is of order
$\log^* n$, the number of black-box calls to $f$ for the simulation of each
$H_{(i,j,\nu)}$ is $O(\log^* n)$. Using these values, along with
Eq.~(\ref{Nscaling}), we obtain the number of black-box queries as in
Eq.~(\ref{calls}).
\end{proof}

Another issue is the number of auxiliary operations, which is the number of
operations that are required due to the overhead
in calculating $\Upsilon(x,i,j)$. It is necessary to perform bit comparisons
between a maximum of $z_n+2$ numbers in the first step, and each has $n$ bits.
This requires $O(n \log^* n)$ operations. In the next step the number of bits is
$O(\log_2 n)$ bits, which does not change the scaling. Hence the number of
auxiliary operations is
\[
O\left(n(\log^*n)^2 d^2 5^{2k}(d^2\tau)^{1+1/2k}/\epsilon^{1/2k}\right).
\]
This scaling is superior to the scaling $n^{10}$ in Ref.\ \cite{aha}.

Next we consider the error introduced by calculating the matrix
elements to finite precision. Given
that the matrix elements are represented by $2n'$ bit integers, the error
cannot exceed $\| H \| /2^{n'}$. The error in calculating $\exp(-iH_{(i,j,\nu)}t)$
will not exceed $\tau/2^{n'}$ \cite{aha}, so the error in the integrator due
to the finite precision does not exceed $4m5^k\tau/2^{n'}$. This error can then
be kept below $\epsilon/2$ by choosing $$n'>5+\log_2(\tau d^2 5^k/\epsilon).$$
The total error may be kept below $\epsilon$ by choosing the integrator such
that the integration error does not exceed $\epsilon/2$.

\section{Conclusions}
We have presented a scheme for simulating sparse Hamiltonians
that improves upon earlier methods in two main ways. First, we have examined
the use of higher order integrators to reduce the scaling to be close to linear
in $\|H\|t$. Second, we have significantly improved the algorithm for the
decomposition of the Hamiltonian, so the scaling of the number of black-box
calls is close to $\log^*n$, rather than polynomial in $n$. In addition we
have shown that the scaling cannot be sublinear in $\|H\|t$ (for reasonable
values of $n$).

\acknowledgements
This project has been supported by the Australian Research Council, the
University of Queensland, and Canada's NSERC, iCORE, CIAR and MITACS.
R.C. thanks Andrew Childs for helpful discussions.

\end{document}